\documentclass[12pt,letterpaper]{article}
\pdfoutput=1

\usepackage{graphicx,array}
\usepackage{color}
\usepackage{latexsym}
\usepackage{amsthm}
\usepackage{amsmath}
\usepackage{amssymb}
\usepackage[numbers,sort&compress]{natbib}
\usepackage{bm}
\usepackage{slashed}
\usepackage{mathrsfs}
\usepackage{hyperref} %Automatically links \label and \ref commands; Always load last
\hypersetup{
    colorlinks=true,       % false: boxed links; true: colored links
    linkcolor=red,          % color of internal links
    citecolor=blue,        % color of links to bibliography
    filecolor=magenta,      % color of file links
    urlcolor=blue           % color of external links
}
\usepackage[all]{hypcap} %Link navagates to top of figure instead of caption (below fig)
\usepackage{subfigure}
\usepackage{multirow}

\usepackage{latexsym}

\newcommand{\nc}{\newcommand}  
\newcommand{\mc}{\mathcal}

\nc{\beq}{\begin{equation}}  
\nc{\eeq}{\end{equation}}  
\nc{\beqa}{\begin{eqnarray}}  
\nc{\eeqa}{\end{eqnarray}}  
\nc{\bit}{\begin{itemize}}  
\nc{\eit}{\end{itemize}}  

\def\TeV{\mathrm{TeV}}     % TeV
\def\GeV{\mathrm{GeV}}     % GeV
\def\ifb{\mathrm{fb}^{-1}} % fb^-1
\def\fb{\mathrm{fb}} % fb

\usepackage{floatrow}
\newfloatcommand{capbtabbox}{table}[][\FBwidth]

\usepackage{blindtext}

\title{\bf Hexapod Coloron at the LHC}
\author{\large Yang Bai$^{a}$, Sida Lu$^{a}$ and Qian-Fei Xiang$^{a,b}$}
\date{\small \it 
$^a$Department of Physics, University of Wisconsin-Madison, Madison, WI 53706, USA  \vspace{1mm}\\
$^b$Key Laboratory of Particle Astrophysics, Institute of High Energy Physics,\\
Chinese Academy of Sciences, Beijing 100049, China
}

\begin{document}

\maketitle

\setlength{\parskip}{0.2ex}

\begin{abstract}
Instead of the usual dijet decay, the coloron may mainly decay into its own ``Higgs bosons", which subsequently decay into many jets. This is a general feature of the renormalizable coloron model, where the corresponding ``Higgs bosons" are a color-octet $\Theta$ and a color-singlet $\phi_I$. In this paper, we perform a detailed collider study for the signature of $pp \rightarrow G' \rightarrow (\Theta \rightarrow gg) (\phi_I \rightarrow gg q\bar{q})$ with the coloron $G'$ as a six-jet resonance. For a light $\phi_I$ below around 0.5 TeV, it may be boosted and behave as a four-prong fat jet. We also develop a jet-substructure-based search strategy to cover this boosted $\phi_I$ case. Independent of whether $\phi_I$ is boosted or not, the 13 TeV LHC with 100 fb$^{-1}$ has great discovery potential for a coloron with the mass sensitivity up to 5 TeV.  
\end{abstract}

\thispagestyle{empty}  
\newpage  
  
\setcounter{page}{1}  

\begingroup
\hypersetup{linkcolor=black}
%\tableofcontents
\endgroup

\newpage

%==================================
% Introduction
%==================================
\section{Introduction}\label{sec:Introduction}
The renormalizable coloron model (ReCoM)~\cite{Bai:2010dj,Bai:2018jsr} is one of simplest gauge extensions of the Standard Model (SM). Due to the coloron's flavor-blind nature and orthogonality to the electroweak sector, its mass could be at the TeV scale and consistent with various experimental constraints. Therefore, the coloron is an ideal particle for the Large Hadron Collider (LHC) to search for. Besides its  phenomenological motivation, the coloron, or its cousin ``axi-gluon", has been predicted in many scenarios beyond the SM~\cite{Hill:1991at,Hill:1993hs,Chivukula:1996yr,Simmons:1996fz,Hall:1985wz,Frampton:1987dn,Bagger:1987fz,Kilic:2009mi} (see also \cite{Dobrescu:2007yp,Sayre:2011ed,Bai:2011mr,Chivukula:2013xka,Chivukula:2014rka,Bai:2017zhj,Draper:2018tmh} for related phenomenological studies). 

At the LHC, the simplest signatures of colorons include dijet and $t\bar{t}$ resonances. Based on narrow dijet resonance searches of the 13 TeV LHC with around 36 fb$^{-1}$ data, both CMS~\cite{CMS:2017xrr} and ATLAS~\cite{Aaboud:2017yvp} collaborations have set impressive constraints on the coloron mass to be above around 6 TeV at 95\% confidence level (CL). The constraints from $t\bar{t}$ resonance searches are weaker~\cite{Aaboud:2018mjh,Sirunyan:2017uhk}. For both existence searches, the coloron is assumed to decay 100\% into two quarks, which may be a too-simplicity assumption. As analyzed in Refs.~\cite{Bai:2010dj,Bai:2018jsr}, additional scalar bosons exist in the spectrum for a perturbative theory. This is a universal behavior of spontaneous breaking of renormalizable and perturbative gauge theory, for which one or more scalar bosons become the company of the massive gauge boson. The famous example is for sure the SM Higgs boson as a bi-product of the Higgs mechanism to explain $W$ and $Z$ gauge boson masses. 

More specifically and in ReCoM, a minimal scalar field, that is bi-fundamental under the $SU(3)_1\times SU(3)_2$ gauge symmetry, develops a non-zero vacuum expectation value (VEV) to break the gauge symmetry to the QCD $SU(3)_c$~\cite{Bai:2017zhj}. There are three scalar bosons in the spectrum, $\phi_R$, $\phi_I$ and $\Theta$, with the first two color-singlet and the last one color-octet. More interestingly, there is spontaneous breaking of some enhanced global symmetries such that both $\phi_I$ and $\Theta$ could be pseudo Nambu-Goldstone Bosons (PNGB's) and naturally lighter than the coloron $G^\prime$. As a result, there are two additional decay channels for the coloron, $G^\prime \rightarrow \Theta \Theta$ and $G^\prime \rightarrow \Theta\phi_I$, which could have decay branching fractions larger than the one of coloron decaying into two quarks. There are two consequences for those additional decay channels. The first one is that the dijet constraints on the coloron masses could be dramatically relaxed and a coloron below 3 TeV can still survive the current experimental searches. The second one is that there are additional opportunities for the LHC to discover the massive coloron via other signatures, which will be the main focus of this paper. 

Depending on how $\Theta$ and $\phi_I$ decays, a high multiplicity of quarks and gluons or jets after hadronization are anticipated in the final state from the coloron's cascade decays. For instance, $\Theta$ can decay into two gluons at one loop or decay into $\phi_I$ plus $q\bar{q}$ at tree-level via an off-shell coloron. The decays of $\phi_I$ are more complicated. For a heavier $\phi_I$ above around 500 GeV, its main decay channel is a four-body one, $\phi_I \rightarrow gg q\bar{q}$, happening at one-loop. For a lighter $\phi_I$, its main decay channels are into two electroweak gauge bosons at three-loop and with a displaced vertex. In Ref.~\cite{Bai:2018jsr}, a throughout list of possible, $\mathcal{O}(50)$, final states has been tabulated. In this paper, rather than making an extensive study of all possible signatures, we mainly concentrate one particular decay chain $pp \rightarrow G^\prime \rightarrow (\Theta \rightarrow gg) (\phi_I \rightarrow q\overline{q}g g)$ and perform a detailed collider simulation to estimate the discovery potential at the 13 TeV with 100 fb$^{-1}$.

\begin{figure}[thb!]
\begin{center}
\includegraphics[width=0.45\textwidth]{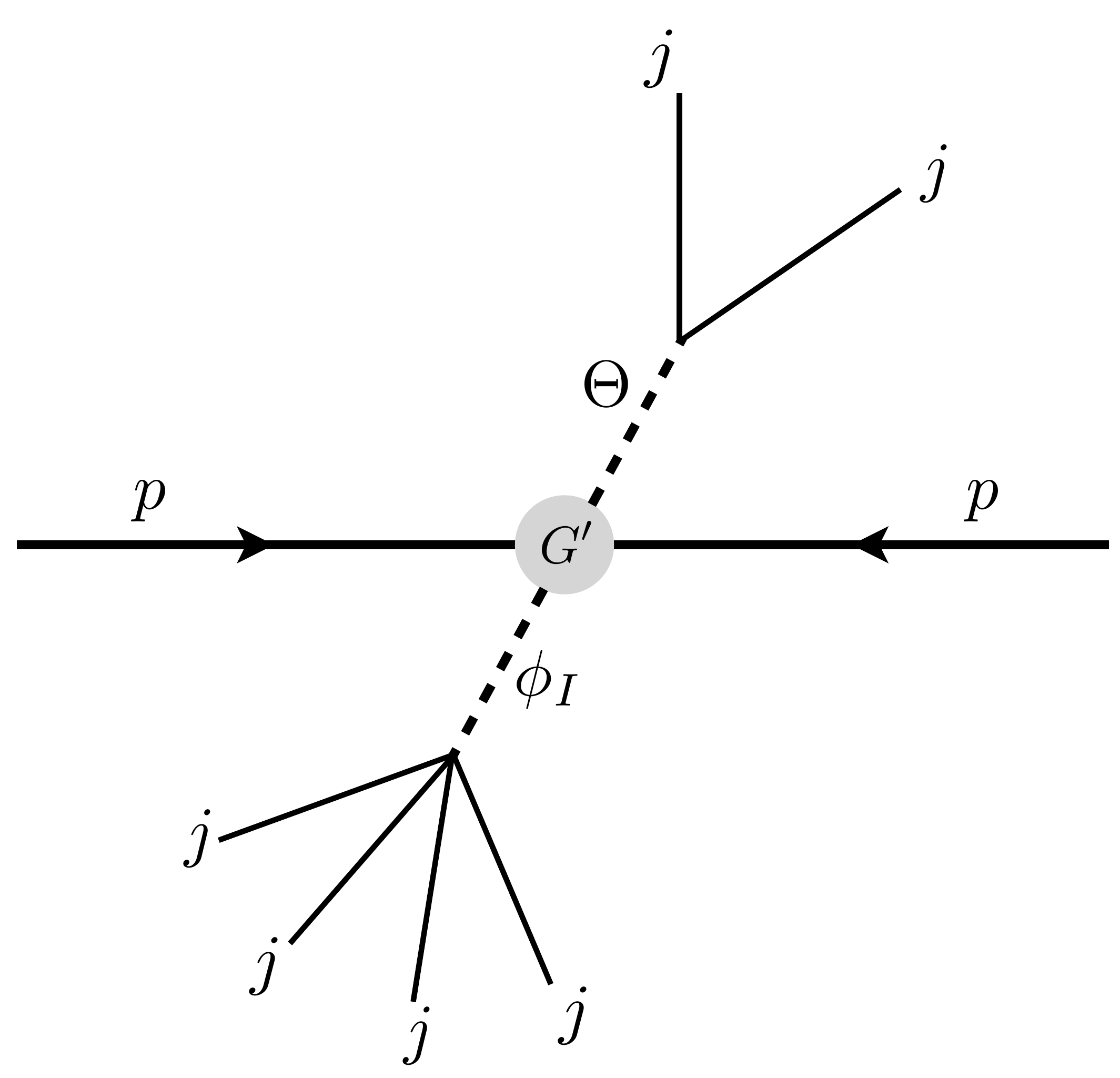}
\hspace{6mm}
\includegraphics[width=0.45\textwidth]{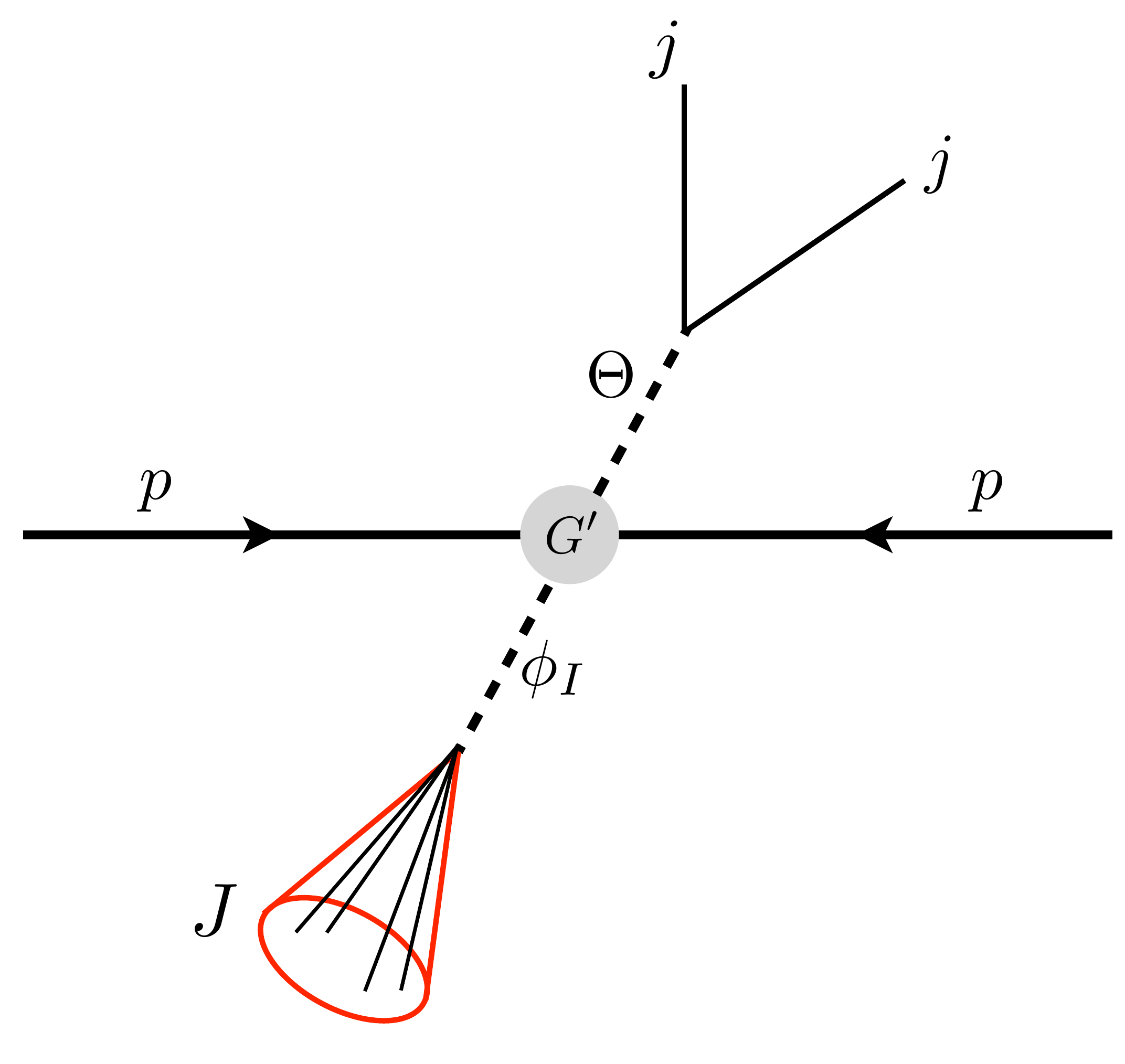}
\vspace*{-0.3cm}
\caption{Left panel: the schematic plot for the coloron signature at the LHC as a six-jet resonance when $M_{\phi_I} \sim M_{G^\prime} - M_\Theta$. Right panel: when $M_{\phi_I} \ll M_{G^\prime} - M_\Theta$, the particle $\phi_I$ is boosted such that it may behave as a four-prong fat jet.}
\label{fig:cartoon}
\end{center}
\end{figure}

Even after we fix the decay chain and depending on the mass spectra, there are still two possible different collider signatures. For the mass spectrum with $M_{\phi_I} \sim M_{G^\prime} - M_\Theta$, the $\phi_I$ scalar from coloron decays is not boosted, so the four partons from $\phi_I$ decays are likely to form four isolated jets if one uses the ordinary jet-finding algorithm. Together with the two jets from $\Theta$ decays, the coloron $G^\prime$ behaves as a {\it six-jet resonance} with the schematic process shown in the left panel of Fig.~\ref{fig:cartoon}. We will develop a bump search strategy for this new type of signature and estimate the LHC sensitivities on the model parameter. The second possibility relies on the PNGB nature of $\phi_I$, which could be much lighter than both $\Theta$ and $G^\prime$. If $M_{\phi_I} \ll M_{G^\prime} - M_\Theta$, the $\phi_I$ particle from coloron decays is likely to be boosted and the four partons from its decay will be collimated. For this case, if one chooses the jet-finding algorithm with a larger value of geometric size, the four partons from $\phi_I$ decays can be grouped into a single fat jet, see the right panel of Fig.~\ref{fig:cartoon}. We will also develop a jet-substructure-based analysis for this type of signatures. 

Our paper is organized as follows. In Section~\ref{sec:model}, we provide a short summary for the ReCoM including the decay branching fractions of all three relevant particles, $G'$, $\Theta$ and $\phi_I$. We show the production times branching fractions for the signatures in Section~\ref{sec:production}. In Section~\ref{sec:6-jet}, we study a few kinematic variables to optimize the searches for the six-jet resonances, while in Section~\ref{sec:fat-jet} we perform a jet-substructure-based analysis for the case with a fat $\phi_I$ jet. We conclude our paper in Section~\ref{sec:conclusion}.

%%%%%%%%%%%%%%%%%%%%%%%%%%%%%%
\section{A short summary of the ReCoM}
\label{sec:model}
%%%%%%%%%%%%%%%%%%%%%%%%%%%%%%
As a simple extension of the SM, the ReCoM serves as a representative one for a class of models with a color-octet massive gauge boson. The ReCoM model extends the SM with an $SU(3)_1\times SU(3)_2 \times SU(2)_W \times U(1)_Y$ gauge group and the following spontaneous breaking of the gauge symmetry
\beq
SU(3)_1\times SU(3)_2  \xrightarrow{\langle\Sigma\rangle}{SU(3)_c}  \,, 
\eeq
with the bi-fundamental scalar field $\Sigma$ transforming as $(3, \overline{3}, 1, 1)$ under the gauge group and developing a VEV from minimizing a potential. The massive gauge bosons, corresponding to the broken gauge generators, are called the ``coloron", $G^\prime$. All SM quarks are charged in the fundamental representation of $SU(3)_1$, such that the coloron couples to quarks in the flavor-blind way and there is no flavor-changing neutral current or stringent constraints from flavor physics. 

Other than the massive coloron field, there are three more scalar particles: a color-octet real scalar field $\Theta^a$ and two scalar singlets $\phi_R$ and $\phi_I$. The masses of  heavy coloron and $\phi_R$ are largely determined by the symmetry breaking scale $f_\Sigma$, while $\Theta$ and $\phi_I$ could be lighter due to their PNGB nature~\cite{Bai:2018jsr}. As emphasized in Ref.~\cite{Bai:2018jsr}, lighter $\Theta^a$ and $\phi_I$ could dramatically modify decays of the coloron, suppress the coloron dijet decay branching fraction and allow a lighter coloron to be searched for at the LHC. If kinematically allowed, the two-body tree-level decay widths of the coloron have~\cite{Bai:2018jsr,Bai:2010dj}
\beqa
\begin{aligned}
&\Gamma(G^\prime \to j\,j)=5\,\Gamma(G^\prime \to t\,\bar{t})=\dfrac{5\,\alpha_s}{6}\tan^2\theta\,M_{G^\prime} ~,\\
%\left(1-\dfrac{4m^2_q}{M^2_{G^\prime}}\right)^{1/2},\\
&\Gamma(G^\prime\to \Theta\Theta)=\dfrac{\alpha_s}{32\tan^2\theta}(1-\tan^2\theta)^2 M_{G^\prime}\left(1-\dfrac{4M^2_\Theta}{M^2_{G^\prime}}\right)^{3/2}   ~,\\
&\Gamma(G^\prime\to \Theta\phi_I)=\dfrac{\alpha_s}{72\tan^2\theta}(1+\tan^2\theta)^2 M_{G^\prime}\left[1-2\dfrac{M^2_\Theta+M^2_{\phi_I}}{M^2_{G^\prime}}+\dfrac{(M^2_\Theta - M^2_{\phi_I})^2}{M^4_{G^\prime}}\right]^{3/2} ~.
\end{aligned}
\label{eq:three-decay-channel}
\eeqa
In the first line of the above equation for the decays into quarks, we have summed the five light flavors and ignored the quark masses. The mixing angle, $\tan{\theta}$, is related to the quark couplings to coloron as $g_s  \tan{\theta}\, \overline{q} \gamma^\mu T^a G^{\prime a}_\mu q$. So, for a small value of $\tan{\theta}$, the branching ratio for the coloron dijet decays is suppressed. Requiring the gauge couplings to be perturbative, the mixing angle is constrained to be $0.15 \lesssim \tan{\theta} \lesssim 6.7$. 

\begin{figure}[thb!]
\begin{center}
\includegraphics[width=0.48\textwidth]{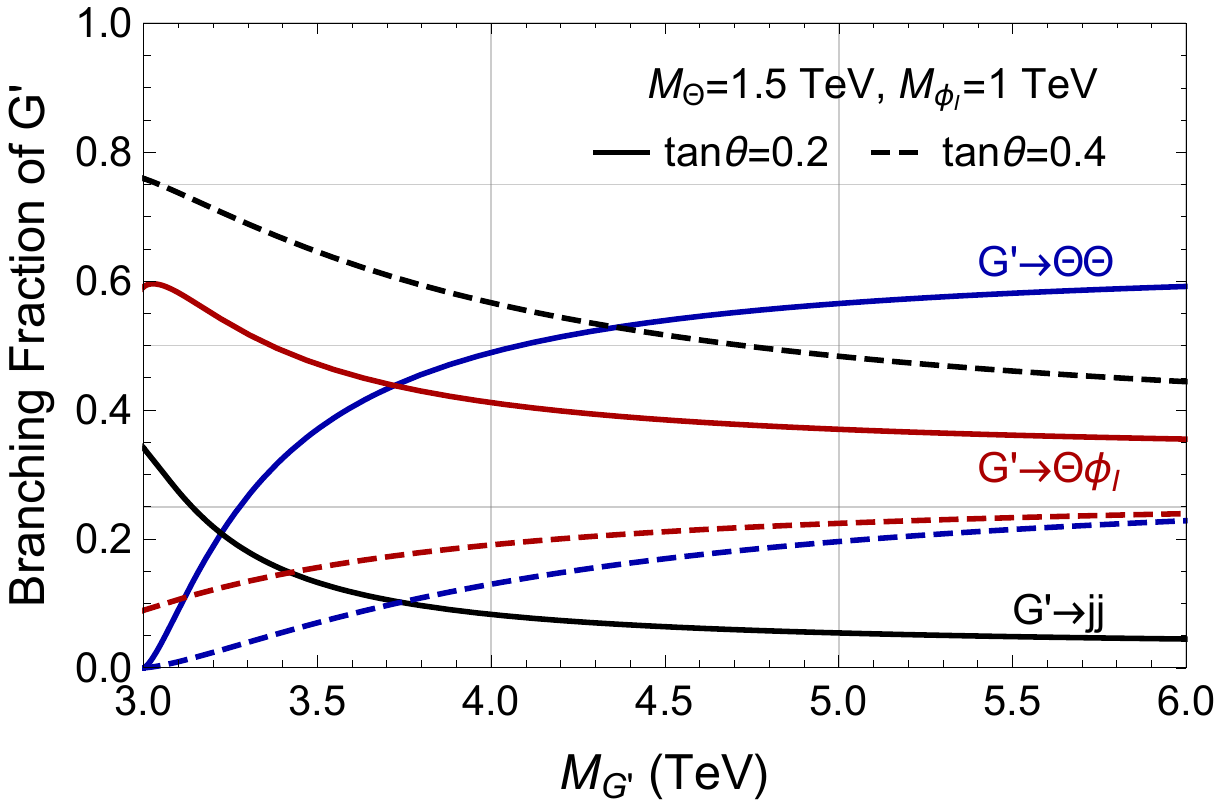}
\hspace{4mm}
\includegraphics[width=0.48\textwidth]{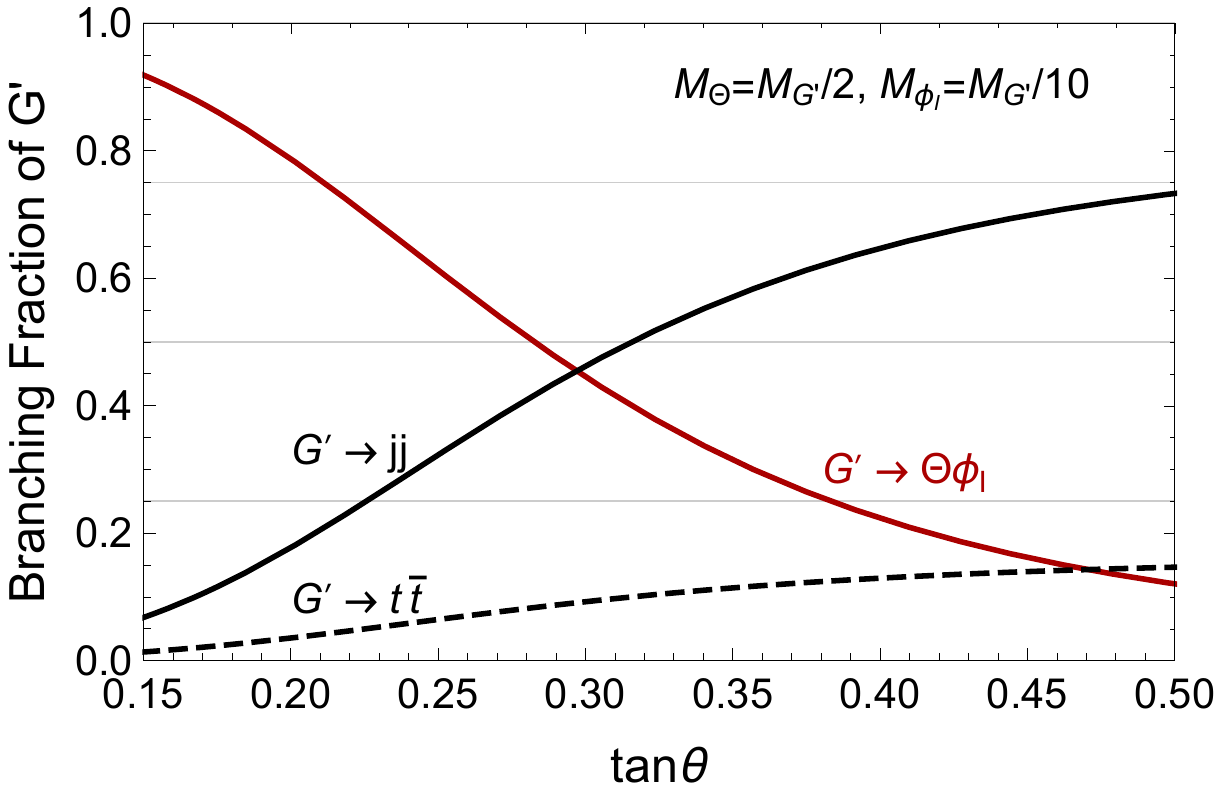}
\vspace*{-0.6cm}
\caption{Branching fraction of $G^\prime$ as a function of $G^\prime$ mass or the mixing angle. In the left panel, the $\Theta$ and $\phi_I$ masses are fixed, while in the right panel their ratios over the coloron mass are fixed. 
%The $j$ in the final state includes $u, d, s, c, b$ quarks. 
The branching fraction of $t\bar{t}$ final state (not shown on the plot) is roughly 1/5 of the $jj$ final state.}
\label{fig:gp_fraction}
\end{center}
\end{figure}

Numerically, we show the branching fractions of coloron decays in Fig.~\ref{fig:gp_fraction}. In the left panel of Fig.~\ref{fig:gp_fraction}, we fix the $\Theta$ and $\phi_I$ masses to be 1.5 TeV and 1 TeV, respectively, while in the right panel we fix the mass ratios to be $M_\Theta/M_{G'} = 1/2$ and $M_{\phi_I}/M_{G'} = 1/10$. For the perturbative-allowed range of $\tan{\theta}$, the ratio of the coloron width  over mass is below 5\%, so the coloron behaves as a narrow resonance at the LHC.

There are two main decay channels for the color-octet scalar $\Theta$: $\Theta\to gg$ and $\Theta \to q\bar{q} \phi_I$. From one-loop diagrams, its decay into two gluons has the form of 
\begin{align}
\Gamma(\Theta\to gg)=\dfrac{15\,\alpha^2_s \, \overline{\mu}^2_\Sigma}{128\,\pi^3 M_\Theta}\left(\dfrac{\pi^2}{9}-1\right)^2(1+r_\mc{A}), \quad \overline{\mu}_\Sigma = \dfrac{3\, g_s (1+\tan^2\theta)}{2\, M_{G^\prime} \tan\theta}\left(M^2_\Theta - \dfrac{8}{9}M^2_{\phi_I}\right) ~,
\end{align}
with the parameter $r_{\cal A}$ given by
\begin{align}
r_\mc{A} = \dfrac{3}{32\sqrt{2}(\pi^2 - 9)}\dfrac{\mc{A}\left[M^2_\Theta/(4M^2_{G^\prime})\right]}{1-8M^2_{\phi_I}/(9M^2_\Theta)} \,, \quad \mc{A}(\tau) = \left[2\tau^2+3\tau+3(2\tau-1)\arcsin^2\sqrt{\tau}\right]\tau^{-2}.
\end{align}
Numerically, $r_{\cal A}\approx 0.53$ for a hierarchic mass spectrum $M_{\phi_I} \ll M_\Theta \ll M_{G'}$. Mediated by an off-shell $G'$, the three-body decay channel $\Theta \to q\bar{q} \phi_I$ has the decay width as
\begin{align}
\Gamma(\Theta \to j j \phi_I)\approx 5 \,\Gamma(\Theta \to \overline{t}t \phi_I)=\dfrac{5\, \alpha^2_s M^5_\Theta}{576\pi M^4_{G^\prime}}(1+\tan^2\theta)^2 \mc{F}(M^2_\Theta / M^2_{G^\prime}, M^2_{\phi_I} / M^2_{G^\prime}) ~,
\end{align}
with the function $\mc{F}(x, y)$ defined in Ref.~\cite{Bai:2018jsr} and having the limited value $\mc{F}(x, y) \xrightarrow{y \ll x \ll 1} 1$. In the limit of $M_{\phi_I} \ll M_\Theta \ll M_{G'}$, the ratio of the two decay channels is approximately 
\beqa
\frac{\Gamma(\Theta \to j j \phi_I)}{\Gamma(\Theta\to gg)} \approx 0.18 \times \left( \frac{\tan{\theta}}{0.3} \right)^2 \left( \frac{M_\Theta/M_{G'}}{1/3} \right)^2 \, . 
\eeqa
So, for a small value of $\tan{\theta}$ and $M_\Theta/M_{G'}$,  the dominant decay channel is mainly $\Theta\to gg$. We will not consider other small decay channels like $\Theta \rightarrow qq\bar{q}\bar{q}$ through two off-shell $G^\prime$'s, which is even more suppressed. 

The decay modes of $\phi_I$ are much more complicated due to its $\mc{P}$-even and $\mc{C}$-odd nature. For a heavier $\phi_I$ above around 500 GeV, the main decay channel is one-loop four-body into $ggq\overline{q}$ with the decay width as
\begin{equation}
\begin{aligned}
&\Gamma(\phi_I \to q\bar{q}gg)=\dfrac{\alpha^5_s}{840(12\pi)^4}(C_\Theta + C^\prime_\Theta)^2 \dfrac{(1+\tan^2\theta)^4}{\tan^2\theta}\dfrac{M^{11}_{\phi_I}}{M^4_\Theta \, M^6_{G^\prime}} ~,
%\\ &C_\Theta \approx 4.61\left(1-0.58\dfrac{M^2_{\phi_I}}{M^2_\Theta}+0.018\dfrac{M^2_\Theta}{M^2_{G^\prime}}\right).
\label{eq:Gamma_phi}
\end{aligned}
\end{equation}
with $C_\Theta \approx 4.6$ for $M_{\phi_I} \ll M_\Theta \ll M_{G'}$ and $C^\prime_\Theta$ expected to be $\mc{O}(1)$. For a lighter $\phi_I$, other two-body three-loop decay channels $\phi_I \rightarrow W^+W^-, ZZ, Z\gamma$ could be dominant, although the lifetime of $\phi_I$ could be long enough to have displaced vertexes at the LHC. Given the uncertain results from the three-loop calculations, we will simply assume that the decay branching fraction of $\phi_I \to q\bar{q}gg$ is close to 100\% in our following analysis. We will also ignore the possibly long lifetime of $\phi_I$ and simply assume that $\phi_I$ decays promptly inside detectors. The long-lived $\phi_I$ case requires its own dedicated studies as in Refs.~\cite{Khachatryan:2016sfv,Aaboud:2017iio}. 

%==================================
% Production and signatures
%==================================
\section{Productions and new signatures at the LHC}
\label{sec:production}
%==================================
At the LHC and for a heavy coloron above around 3 TeV, the main production is a single resonance production. Using the narrow-width approximation and multiplying the cross section by the $K$-factor of 1.2~\cite{Chivukula:2013xla}, the $G'$ production cross section is approximately $(20,1.8, 0.20,0.023)$~pb for $M_{G'}=2,3,4,5$~TeV for $\tan{\theta}=1.0$, using the MSTW parton distribution function~\cite{Martin:2009iq}.

For the three main decay channels in \eqref{eq:three-decay-channel}, the simplest signature to search for the coloron is the traditional one of a dijet resonance. Indeed, both ATLAS and CMS collaborations have set upper constraints on the production cross section times the dijet branching fraction~\cite{CMS:2017xrr,Aaboud:2017yvp}. Using the signal acceptance of around 0.6~\cite{CMS:2017xrr}, we will recast their constraints on coloron model parameter space based on the 36~fb$^{-1}$ data. We will also make a simple projection of their sensitivity at 100~fb$^{-1}$. For a small value of $\tan{\theta}$ around 0.2, the branching fraction into dijet can be as small as 5\%, so a weaker constraint on the coloron mass, $M_{G^\prime} \gtrsim 2$~TeV.

The next type of signature is $pp \rightarrow G^\prime \rightarrow (\Theta \rightarrow gg) (\Theta \rightarrow gg)$, with a pair of dijet resonances that form a four-jet resonance. Depending on the $\Theta$ particle mass, its pair-production from QCD interactions may also be important. The possible three-body decay channel, $\Theta \rightarrow \phi_I q \overline{q}$, can generate additional signatures with a larger multiplicity of jets. On other hand, due to the small branching fraction, at least for the benchmark spectrum considered later, we don't consider the three-body decays of $\Theta$ in this paper. There are no dedicated experimental searches for a four-jet resonance with subsequent dijet resonances. For  the searches for a pair of dijet resonances without a four-jet resonance~\cite{Khachatryan:2014lpa,ATLAS-CONF-2017-025},  we have checked and found that they do not constrain the ReCoM parameter space considered later.

The third type of signature has the process of $pp \rightarrow G^\prime \rightarrow (\Theta \rightarrow gg) (\phi_I \rightarrow q\overline{q}g g )$, which will be the focus of this paper. For a heavier $\phi_I$ above around 1 TeV, the four partons from its decay are likely to form four jets, so the coloron $G^\prime$ behaves as a {\it six-jet resonance}. The schematic signature at the LHC is shown in the left panel of Fig.~\ref{fig:cartoon}. On the other hand, if $\phi_I$ is significantly lighter than the mass difference of $G^\prime$ and $\Theta$ or $M_{\phi_I} \ll M_{G^\prime} - M_\Theta$, the $\phi_I$ from the coloron decay could be boosted such that the four partons from its decay may be grouped into a single fat jet. For this case, we will have the signature of $pp \rightarrow G^\prime \rightarrow (\Theta \rightarrow gg) J_{\phi_I}$, with $J_{\phi_I}$ as a four-prong fat jet. The corresponding schematic signature at the LHC is shown in the right panel of Fig.~\ref{fig:cartoon}. 

\begin{figure}[thb!]
\begin{center}
\includegraphics[width=0.5\textwidth]{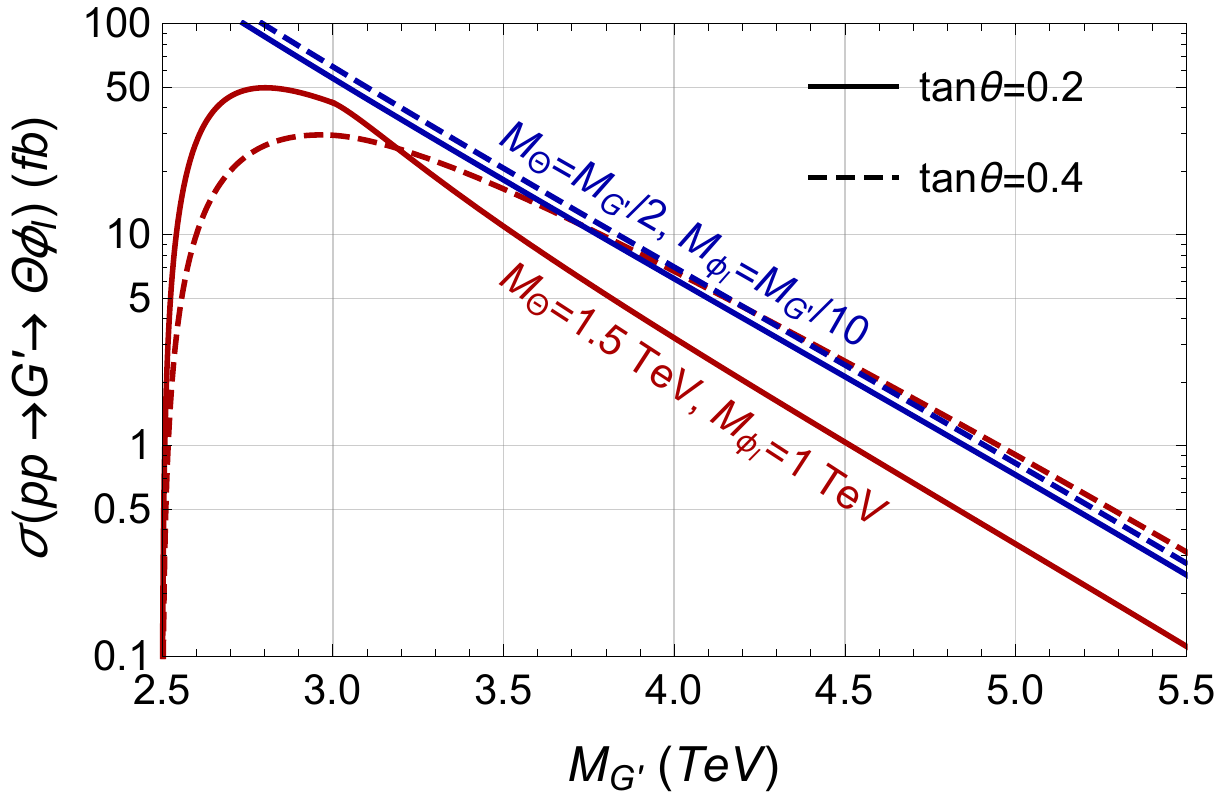}
\vspace*{-0.3cm}
\caption{Production cross section times branching ratio of $pp \rightarrow G' \rightarrow \Theta \phi_I$ at the 13 TeV LHC.}
\label{fig:production}
\end{center}
\end{figure}

To develop detailed search strategies to search for a coloron as a multi-jet resonance,  we introduce two benchmark spectra. For the first one, we fix $M_\Theta=1.5$~TeV and $M_{\phi_I}=1$~TeV, which represents the left panel of Fig.~\ref{fig:cartoon}. For the second one, we choose a hierarchic spectrum with $M_\Theta/M_{G^\prime}=1/2$ and $M_{\phi_I}/M_{G^\prime}=1/10$, for which the $\phi_I$ is likely boosted and matches to the situation in the right panel of Fig.~\ref{fig:cartoon}. At the 13 TeV LHC, we show the corresponding production cross section times branching ratio of $p p \rightarrow  G^\prime \rightarrow \Theta \phi_I$ for different coloron masses in Fig.~\ref{fig:production}. Fixing $M_{G^\prime}=4$~TeV and $\tan{\theta}=0.25$, we show the model information for the two benchmark points (BP's) in Table~\ref{tab:bmps}.

\begin{table}[bth!]
\renewcommand{\arraystretch}{1.2}
\begin{tabular}{c|c|c}
\hline\hline
      & BP1  & BP2\\
  \hline
($M_{G'}$  , $M_\Theta$ , $M_{\phi_I}$, $\mathrm{tan}\theta$)  &    (4, 1.5, 1) TeV    & (4, 2, 0.4) TeV \\
$\mathrm{tan}\theta$  &    0.25   & 0.25 \\
 \hline
  $B({G'} \to \Theta\Theta)$ \quad   $B(G' \to \Theta\phi_I)$ & 41.5\%  \quad 38.1\%  &  0  \quad  61.3\% \\ 
    $B({G'} \to jj)$ \quad   $B(G' \to t\bar{t})$     &     17.0\%  \quad  3.4\%     &   32.3\%  \quad  6.4\%   \\
 $\Gamma_{\rm tot}(G')$     & 98 GeV    &   52 GeV     \\    \hline                                                
 $B(\Theta \to gg)$    & 99.1\%                                                          & 86.5\%  \\
 $B(\Theta \to q\bar{q}\phi_I)$   & 0.9\%                                            & 13.5\%   \\
 \hline 
 $B(\phi \to q\bar{q}gg)$ & 100\%      & 100\% \\
   \hline  
 $\sigma(pp \rightarrow G^\prime\rightarrow \Theta \Theta)$ & 5.1 fb      & 0 fb  \\
  \hline  
 $\sigma(pp \rightarrow G^\prime\rightarrow \Theta \phi_I)$ & 4.7 fb      & 7.5 fb  \\
  \hline 
 \hline
\end{tabular}}
{%
  \caption{Two model benchmark points with some properties of particle decay branching fractions and production cross sections at the 13 TeV LHC.}%
  \label{tab:bmps}
\end{table}
%%%%%%%%%%%%%%%%%%%%%%%%%%%%%%
\section{Search strategies and estimated sensitivities}
\label{sec:strategies}
%%%%%%%%%%%%%%%%%%%%%%%%%%%%%%
In this section, we will develop two independent search strategies for the two types of signatures shown in Fig.~\ref{fig:cartoon}. For both searches, we will try to reconstruct the coloron resonances. For the second case, we will also try to reconstruct the invariant mass of $\Theta$ and the jet mass associated with the $J_{\phi_I}$ fat jet. Before we dive into the individual cases, we first briefly mention our simulation procedures. 

The dominant SM background for the signature we are searching for is the QCD multijet production. The other backgrounds including weak gauge bosons plus jets and top quarks are shown to be subdominant~\cite{Sirunyan:2017anm}. To simulate the background we generate the process $pp\to jj+jjj+jjjj$ at parton level with \texttt{MadGraph~5}~\cite{Alwall:2014hca}. The parton shower, particle decay, and hadronization processes are handled by \texttt{PYTHIA~6}~\cite{Sjostrand:2006za}. The MLM scheme~\cite{Mangano:2006rw} is employed to handle the matching between matrix element and parton shower calculations. \texttt{Delphes~3}~\cite{deFavereau:2013fsa} is utilized to carry out a fast detection simulation with the CMS setup. Jets are clustered using the anti-$k_\mathrm{T}$ algorithm~\cite{Cacciari:2008gp} with a radius parameter of $R = 0.5$, and are accepted only when they have $\lvert \eta \rvert < 2.5$ and $p_T > 50~\GeV$. We also impose a multi-jet trigger requirement with $H_T > 800$~GeV with $H_T$ defined as a scalar summation of all jet $p_T$'s~\cite{Khachatryan:2016bia}. To justify our background simulation, we have also checked the corresponding kinematic distributions against the multi-jet searches for microscopic black holes~\cite{Sirunyan:2017anm,Sirunyan:2018xwt} and found good agreement. The multi-jet search results have also been applied to our model and do not constrain the model parameter space.

For the signal events and other than using \texttt{FeynRules}~\cite{Alloul:2013bka} to implement the ReCoM model, the other Monte Carlo simulation tools are largely the same as background simulation. The only difference is that we use \texttt{PYTHIA~8} \cite{Sjostrand:2014zea} instead of \texttt{PYTHIA~6} such that the four-body decay of $\phi_I$ can be handled. Even though that we will concentrate on the $pp \rightarrow G' \rightarrow \Theta\phi_I$ channel to generate six partons in the final state, we also include the possible signal contaminations of $pp\to\Theta\Theta$ from both $G^\prime$-mediated and QCD interactions. 

%%%%%%%%%%%%%%%%%%%%%%%%%%%%%%
\subsection{Six-jet resonance search for coloron}
\label{sec:6-jet}
%%%%%%%%%%%%%%%%%%%%%%%%%%%%%%
For the signatures in the left panel of Fig.~\ref{fig:cartoon}, the coloron decays into six partons, which could become six isolated jets after parton shower, hadronization and ordinary jet reconstructions. Given the possibility of initial and final state radiations, our general search strategy is very simple: a ``bump" search of resonances composed of six or more energetic jets. One may also hope to construct the intermediate $\Theta$ and $\phi_I$ resonances to further reduce the QCD backgrounds. Given the large combinatorial of $15$, the reconstructions of $\Theta$ and $\phi_I$ are not efficient and will not be considered here. 

\begin{figure}[thb!]
\begin{center}
\includegraphics[width=0.45\textwidth]{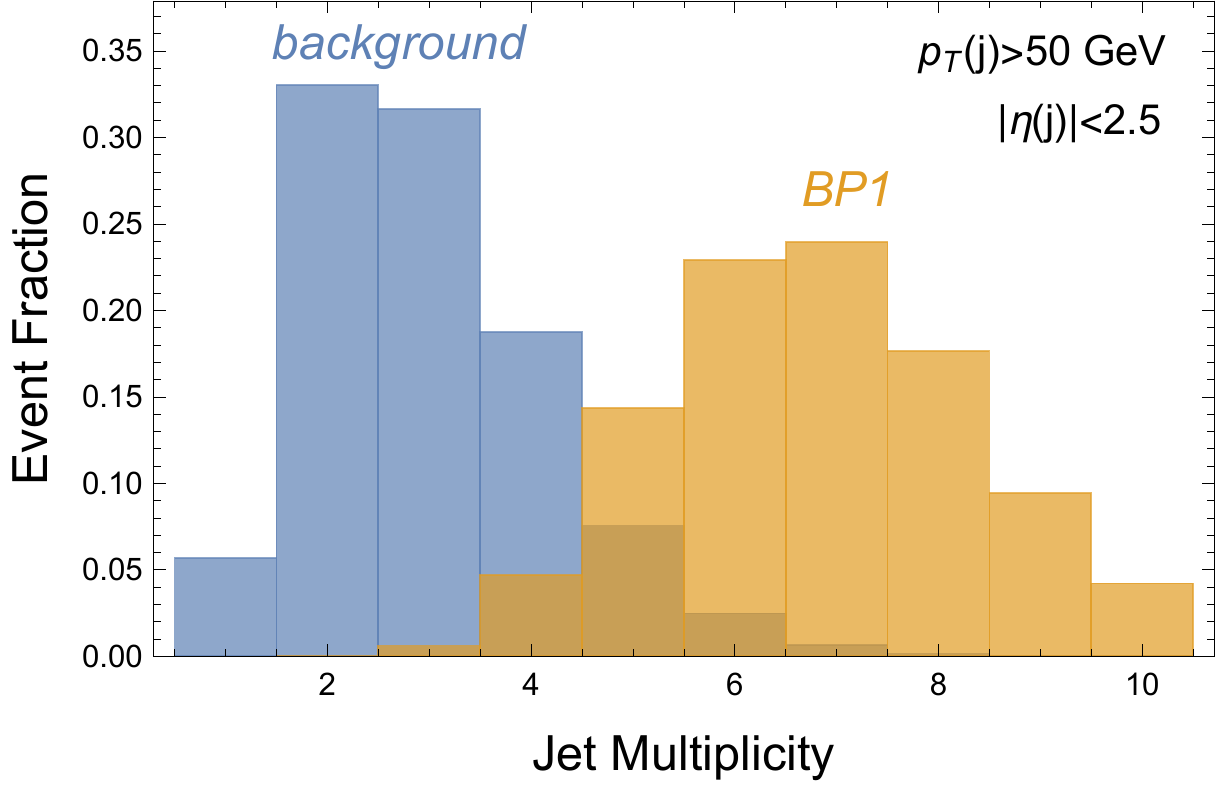}
\hspace{6mm}
\includegraphics[width=0.45\textwidth]{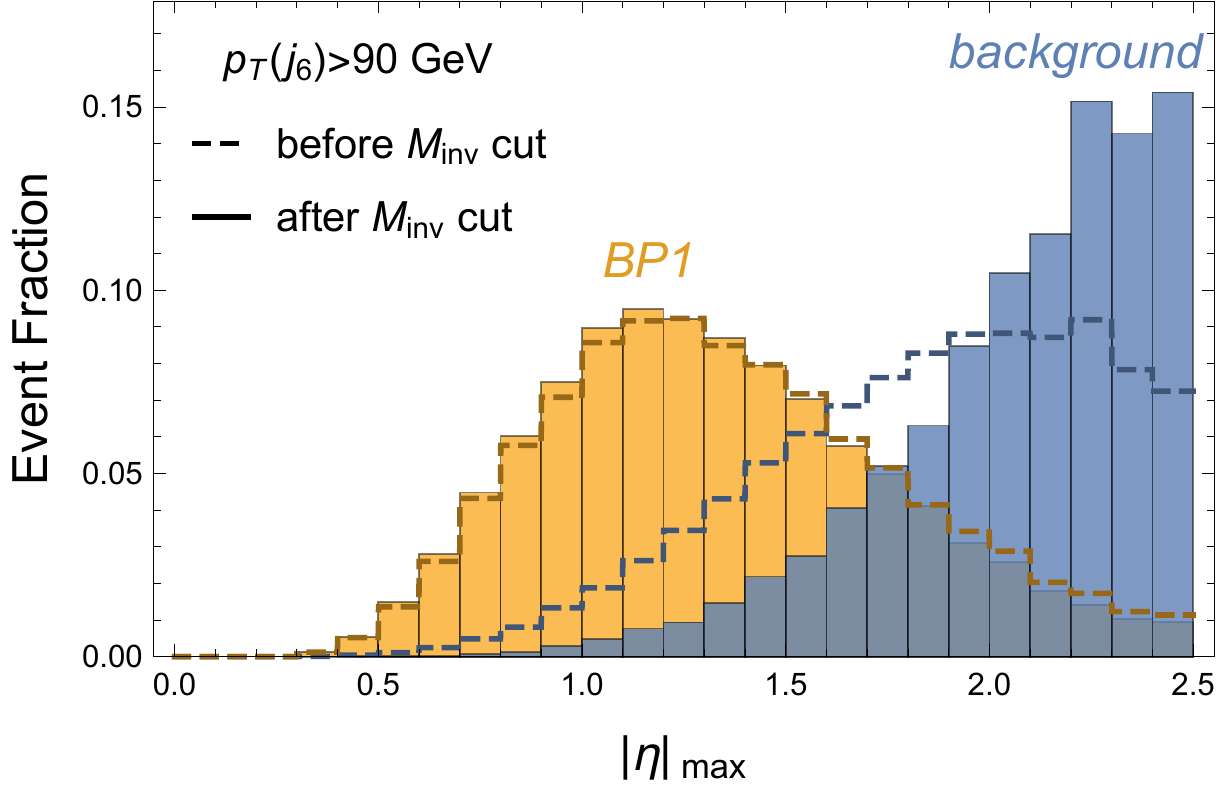}
\vspace*{-0.3cm}
\caption{Left panel: normalized jet-multiplicity distribution for the QCD background and the benchmark model point BP1. Right panel: normalized distributions for $|\eta|_{\rm max}$ (the maximum $|\eta|$ for leading six jets) with and without the cuts on the total invariant mass $\lvert M_{\rm inv} - M_{G^\prime} \rvert / M_{G^\prime} < 15\%$. }
\label{fig:jet-multiplicity-eta}
\end{center}
\end{figure}

There are three useful variables to increase the signal significance: the number of jets $N_j$, $|\eta|$ for jet selection and the invariant mass $M_{\rm inv}$ of all jets. For the first variable $N_j$, we show the normalized distributions of jet multiplicities for both QCD background and the signal BP1 in the left panel of Fig.~\ref{fig:jet-multiplicity-eta}. It is clear that the QCD background has the jet multiplicity distribution peaked at two or three, while the BP1 signal events have a peak at around six and seven. So, imposing a cut on the jet multiplicity with $N_j \geq 6$ can reduce the background and increase the BP1 signal sensitivity. 

Similar to the searches for quark compositeness and because of the Rutherforld scattering in QCD, requiring the jets to be more central or with a small value of $|\eta|$ can improve the signal sensitivity~\cite{Sirunyan:2018wcm,Aaboud:2017yvp}. In the right panel of Fig.~\ref{fig:jet-multiplicity-eta}, we show the variable $|\eta|_{\rm max}$ (the maximum value of $|\eta|$'s of the leading six jets) distributions for both signal and background with or without the additional invariant mass cut $\lvert M_{\rm inv} - M_{G^\prime} \rvert / M_{G^\prime} < 15\%$. It is clear that after imposing the invariant mass cut, the QCD background has a small fraction of events with $|\eta|_{\rm max}$ below around 1.0. So, requiring an upper bound on $|\eta|$ for all jets can efficiently increase the discovery sensitivity.

\begin{figure}[thb!]
\begin{center}
\includegraphics[width=0.5\textwidth]{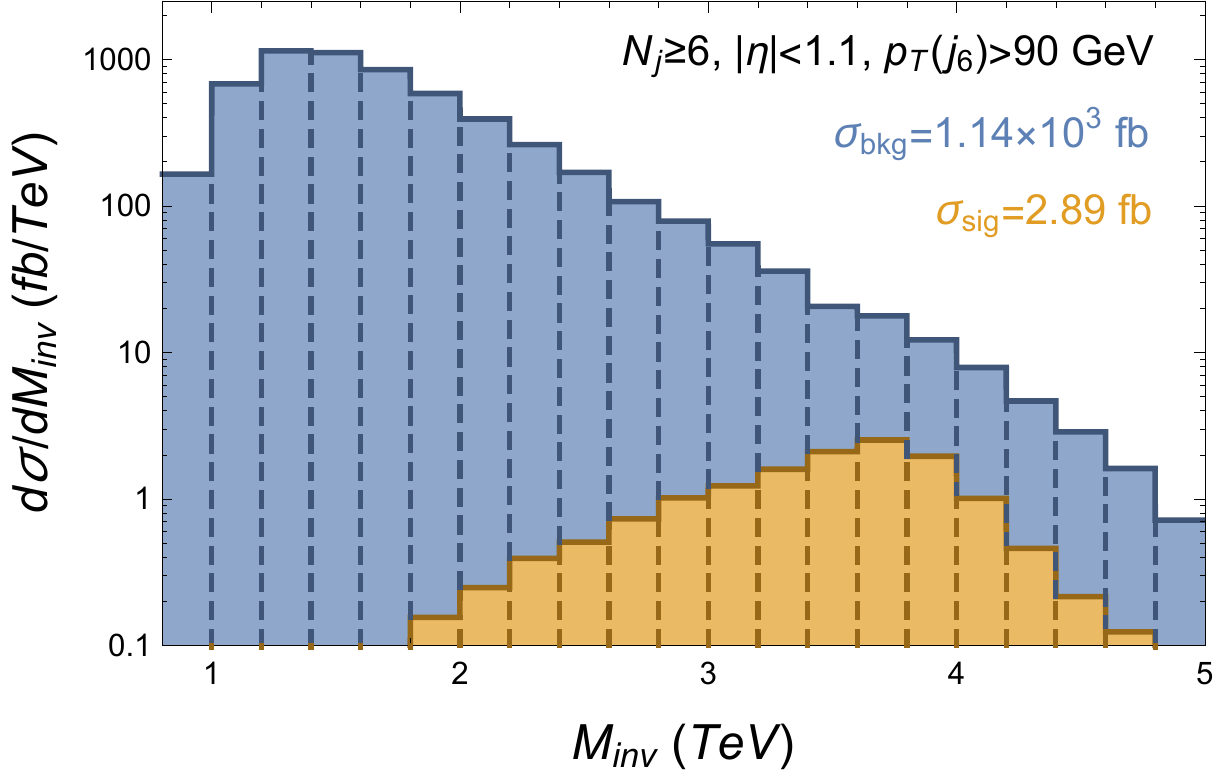}
\vspace*{-0.3cm}
\caption{The signal and background invariant mass distributions for the number of jets equal to or above six. The cuts on other kinematic variables in \eqref{eq:cuts-BP1} have also been imposed. For this BP1 signal, $M_{G'}=4$~TeV.
}
\label{fig:BP1-M_inv}
\end{center}
\end{figure}

For the invariant mass distributions, we show both the background and the signal differential cross section distributions in Fig.~\ref{fig:BP1-M_inv}, after imposing the optimized cuts on other variables. To take into account of the final state radiation jets, the invariant mass variable, $M_{\rm inv}$, includes all jets with $p_T > 50$~GeV. For the signal distributions, one can see that the reconstructed distribution is peaked at around but slightly below the coloron mass, which is due to both energy resolutions and missing jets in the more forward directions. Due to the detector energy resolution, the width of the peak is around 15\% of the coloron mass, so we will choose the mass window cut to be $\lvert M_{\rm inv} - M_{G^\prime} \rvert / M_{G^\prime} < 15\%$. For the background distribution, there is a peak at around 1.4 TeV, which is due to the trigger requirement of $H_T > 800$~GeV and a simulation cut of requiring the invariant mass of partons above 1 TeV.

All together, we have the following optimized cuts for the BP1
\beqa
N_j \geqslant 6 \,, \quad  \lvert \eta \rvert < 1.1 \,, \quad p_T(j_6) > 90~\GeV \,, \quad  \lvert M_{\rm inv} - M_{G^\prime} \rvert / M_{G^\prime} < 15\%  \,.
\label{eq:cuts-BP1}
\eeqa
To demonstrate the efficiencies of different cuts, we show the cut flow of the background and signal events in Table.~\ref{tab:cutflow-BP1}. After imposing all the cuts in \eqref{eq:cuts-BP1}, the signal significance for the BP1 has $S/\sqrt{B} \approx 4.5$ at the 13 TeV LHC with 100 fb$^{-1}$, which definitely shows high discovery potential at the LHC. In Table~\ref{tab:th_vs_ph}, we show the contributions to signal strength from different subprocesses of the BP1. One can see that even though our search strategy for the six-jet resonance is designed for the $\Theta \phi_I$ signal, it can also cover the signal of $pp \rightarrow \Theta \Theta$ with $\Theta$ mainly decaying into two gluons together with additional final-state radiations.

\begin{table}[bht!]
\centering
\renewcommand{\arraystretch}{1.2}
\setlength\tabcolsep{0.4em}
\begin{tabular}{c|c|ccc|ccc}
\hline\hline
\multirow{2}{*}{Cut-flow} & Background             & \multicolumn{3}{c|}{BP1}   & \multicolumn{3}{c}{BP2}\\
& $\sigma~(\fb)$ &$\sigma~(\fb)$ & $\epsilon$ & $S/\sqrt{B}$    &$\sigma~(\fb)$ & $\epsilon$ & $S/\sqrt{B}$\\
\hline  
$N_j \geqslant 6$,  $|\eta| < 2.5$                     & $1.37\times 10^5$   			& 8.72	& 80.6\%	&0.24		& 3.81	& 50.5\%	& 0.10\\
$N_j \geqslant 6$, $\lvert \eta \rvert < 1.1$    & $1.02\times 10^4$   				& 5.09	& 47.1\%	&0.50		& 1.74	& 23.0\%	& 0.17\\
$p_T (j_6)> 90~\GeV$ & $1.14\times 10^3$       				& 2.89	& 26.7\%	&0.86		& 0.75	& 10.0\%	& 0.22\\
$\lvert M_{\rm inv} - M_{G^\prime} \rvert / M_{G^\prime} < 15\%$ &13.14	& 1.62 	& 15.0\% 	&4.5		& 0.47 	& 6.2\%	& 1.3\\
\hline\hline
\end{tabular}
\caption{Cut flow of signal and background cross sections and acceptances. Also shown are the signal significances with an integrated luminosity of $100~\ifb$.}
\label{tab:cutflow-BP1}
\end{table}

\begin{table}[htb!]
\centering
\renewcommand{\arraystretch}{1.2}
\setlength\tabcolsep{0.8em}
\begin{tabular}{c|c|c|c|c|c}
\hline\hline
\multicolumn{2}{c|}{$pp\to G^\prime \to \Theta\phi_I$}  & \multicolumn{2}{c|}{$pp\to G^\prime \to \Theta\Theta$} & \multicolumn{2}{c}{$pp\to \Theta\Theta$ (QCD)}\\ 
\hline
$\sigma~(\fb)$ & $\epsilon$ & $\sigma~(\fb)$ & $\epsilon$ & $\sigma~(\fb)$ & $\epsilon$ \\
\hline
0.90 & 19.3\% & 0.66 & 12.9\% & 0.07 & 6.2\% \\
\hline
\hline  
\end{tabular}
\caption{Contributions to the signal strength of the BP1 from different subprocesses.}
\label{tab:th_vs_ph}
\end{table}

In Table~\ref{tab:cutflow-BP1}, we also show the acceptances and the signal significance for the BP2. Compared to the BP1, the signal acceptance is smaller by a factor of more than two. So, the search strategy developed in this section may not be an optimized one for the BP2. A simple reason for this is that the $\phi_I$ boson is boosted such that it only appears as one high $p_T$ jet. The requirement of having at least six jets with $p_T$ above 90 GeV reduces the BP2 signal strength. In the section~\ref{sec:fat-jet}, we will develop a better fat-jet-based search strategy for the BP2. 

\begin{figure}[bth!]
\begin{center}
\includegraphics[width=0.6\textwidth]{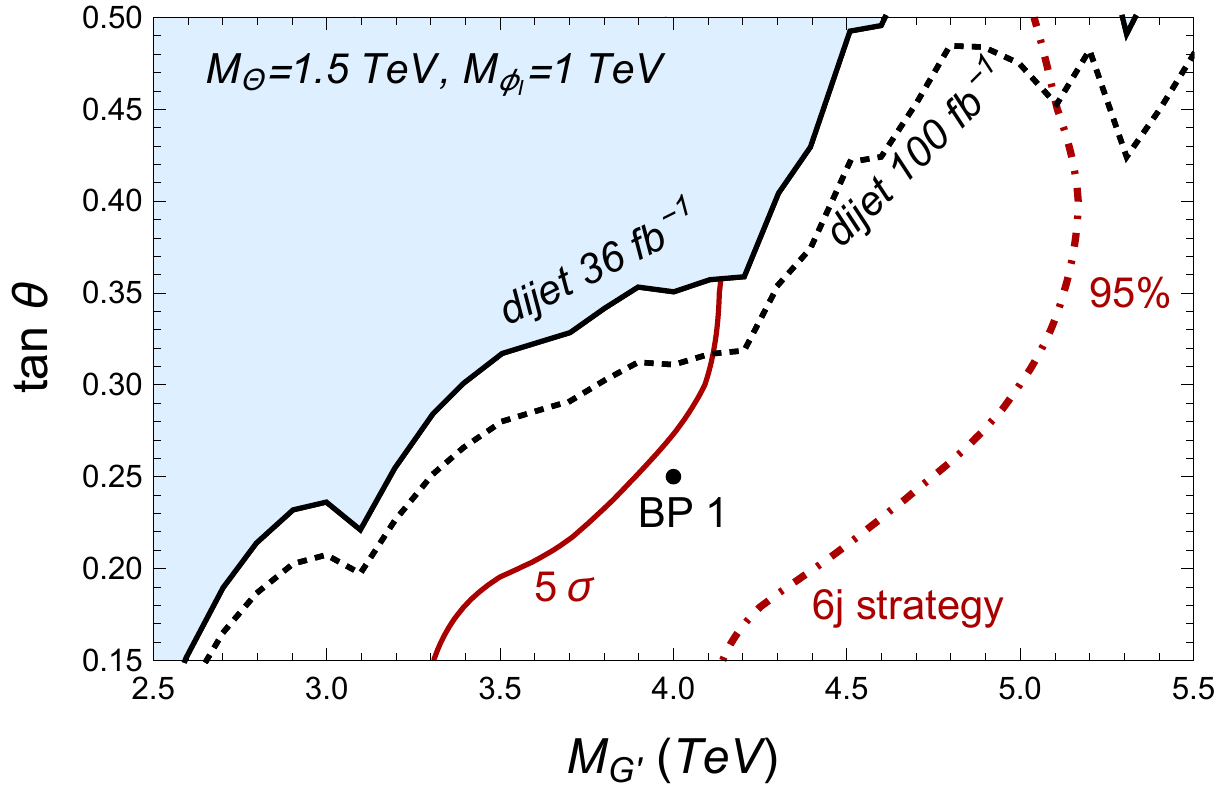}
\vspace*{-0.3cm}
\caption{Estimated sensitivity for different $M_{G^\prime}$ and $\tan\theta$ and fixed $M_\Theta=1.5$~TeV and $M_{\phi_I}=1~\TeV$ at the 13 TeV with 100 fb$^{-1}$, using the cuts in \eqref{eq:cuts-BP1}. The current and projected 95\% CL constraints from the dijet resonance searches are also shown in the black solid and dotted lines~\cite{CMS:2017xrr}.}
\label{fig:BP1_contour}
\end{center}
\end{figure}

Using the cuts in \eqref{eq:cuts-BP1}, we scan the parameter space of $M_{G^\prime}$ and $\tan\theta$ for fixed $M_\Theta = 1.5$~TeV and $M_{\phi_I}=1$~TeV, estimate the discovery sensitivities and show both $5\sigma$ discovery and $95\%$ CL exclusion limits in Fig.~\ref{fig:BP1_contour}. Also shown in this figure are the 95\% CL constraints from dijet resonance searches~\cite{CMS:2017xrr} with 36 fb$^{-1}$ luminosity and the simple projected sensitivity at 100 fb$^{-1}$. A signal acceptance factor of 0.6 is used to obtain the limits in the black solid and dashed lines. Comparing the dijet and six-jet resonance sensitivities, it is clear from this figure that a wide range of parameter space of ReCoM may have $5\sigma$ discovery potential based on searches for a six-jet resonance. 

%%%%%%%%%%%%%%%%%%%%%%%%%%%%%%
\subsection{Coloron as a resonance containing a fat jet $J_{\phi_I}$}
\label{sec:fat-jet}
%%%%%%%%%%%%%%%%%%%%%%%%%%%%%%
For the second type of spectra with $M_{\phi_I} \ll M_{G'} - M_\Theta$, the four partons from $\phi_I$ decays are likely to be collimated and form a single fat-jet $J_{\phi_I}$ (see the right panel of Fig.~\ref{fig:cartoon} for the schematic illustration of this type of signals at the LHC). Meanwhile, the two gluons from $\Theta$ decays will behave as ordinary jets. Therefore, we implement a two-step jet reconstruction method to analyze the events in order to reveal the above features. For all the proto-jets from parton shower, we first use the anti-$k_T$ jet-finding algorithm~\cite{Cacciari:2008gp} with $R=1.0$ to identify the possible $J_{\phi_I}$. More specifically, we choose the $R=1.0$ jet with the largest jet mass $M_J$ as $J_{\phi_I}$. We then remove the components inside $J_{\phi_I}$ from the list and re-cluster the remaining proto-jets into ordinary jets using the $R=0.5$ anti-$k_T$ algorithm. Both large-$R$ jets and small-$R$ jets are required to have $|\eta| < 2.5$.

To optimize the sensitivity for the second type of spectra with a fat $\phi_I$ jet, we have found four useful variables, the fat jet mass $M_J$, an N-subjettiness variable $\tau_2/\tau_1$~\cite{Thaler:2010tr}, the invariant mass of non-fat jet $M_{\rm inv}(j)\equiv M[\sum p_\mu(j)]$ (\textit{all} ordinary jets are included) and the total invariant mass of the fat jet and ordinary jets $M_{\rm inv}(J, j) \equiv M[p_\mu(J) + \sum p_\mu(j)]$. All of those variables except $\tau_2/\tau_1$ are intuitively understood. Naively, since there are four partons from $\phi_I$ decays, one may wonder about another variable $\tau_4/\tau_3$, which is more suitable for a four-prong jet. The decay of $\phi_I \rightarrow gg q\bar{q}$, however, is through off-shell particles rather than on-shell intermediate particles. As a result, the parton energies are more hierarchical, which makes $\tau_4/\tau_3$ less useful to suppress QCD background (see Appendix~\ref{appendix:tau43} for detail).

\begin{figure}[thb!]
\begin{center}
\includegraphics[width=0.45\textwidth]{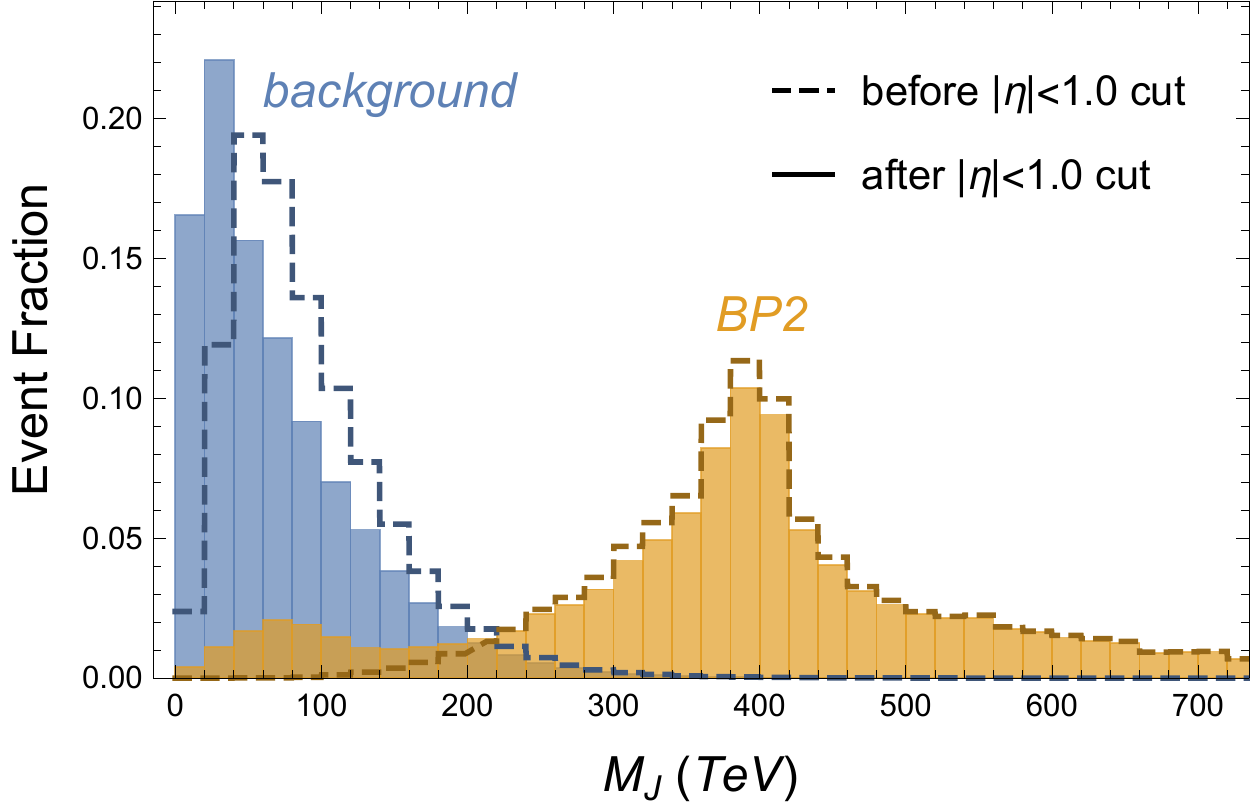}
\hspace{6mm}
\includegraphics[width=0.45\textwidth]{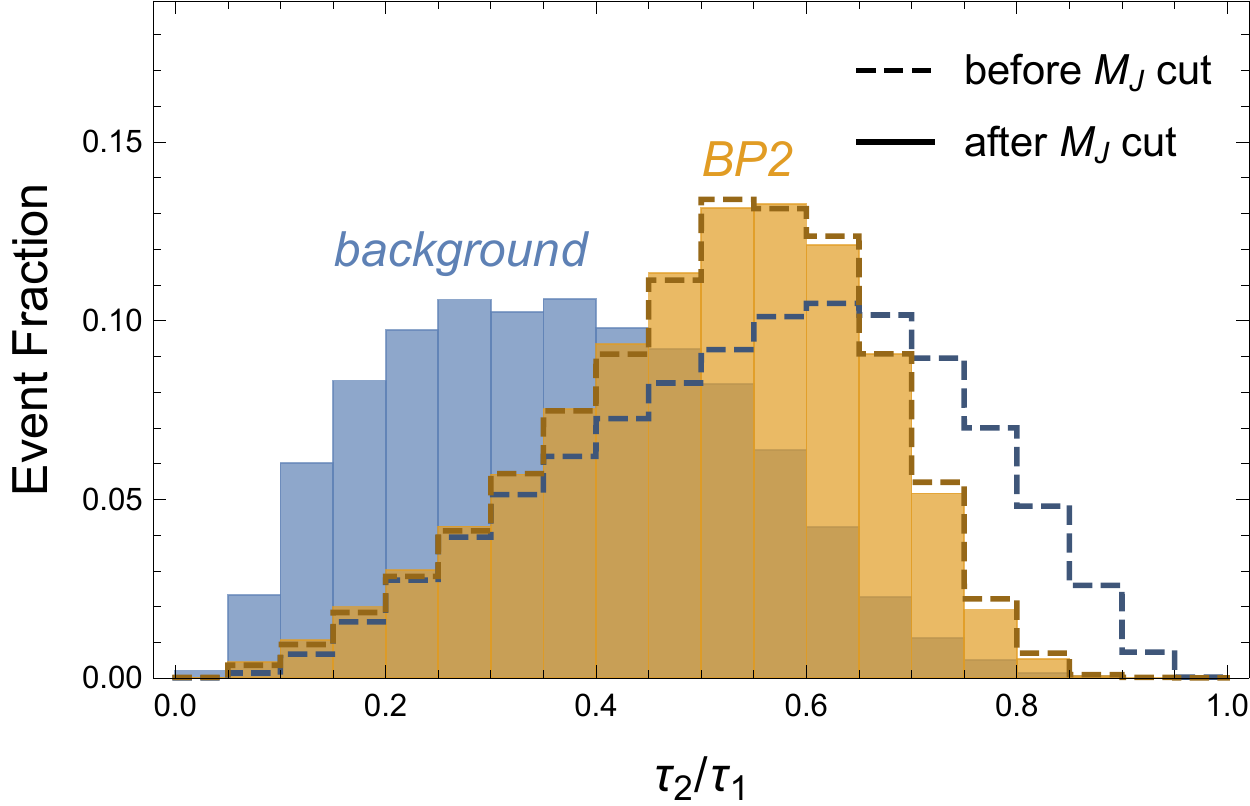}
\vspace*{-0.3cm}
\caption{Left panel: the normalized leading fat-jet mass distributions for the benchmark model point BP2 and QCD background with or without the $|\eta|<1.0$ cut. The anti-$k_T$ jet-finding algorithm with $R=1.0$ has been used. Right panel: normalized $\tau_2/\tau_1$ distributions of the most massive jet for the signal and background with or without the fat-jet mass cut in \eqref{eq:cuts-BP2}.}
\label{fig:jet-mass-tau}
\end{center}
\end{figure}

In the left panel of Fig.~\ref{fig:jet-mass-tau}, we show the normalized fat jet mass distributions for both BP2 and the QCD background. Among more than one possible fat jets, we take the one with the largest jet mass. It is clear from this plot that imposing a cut on the fat-jet mass above around 300 GeV can effectively increase $S/\sqrt{B}$ for the BP2. For the signal events, sometimes the truth $J_{\phi_I}$ may not pass the $|\eta| < 1.0$ cut, so a non-$\phi_I$ jet is chosen as the leading fat-jet candidate, which may have a smaller jet mass and show up in the left part of the signal distribution. In the right panel of Fig.~\ref{fig:jet-mass-tau}, we show the normalized $\tau_2/\tau_1$ distributions for the signal and background. Before we impose the $M_J > 0.8 M_{\phi_I}$ cut, the QCD background is peaked at higher values of $\tau_2/\tau_1$ than the signal. After the $M_J$ cut, the QCD background has a peak at lower values of $\tau_2/\tau_1$, which means that the two-prong feature of the QCD background becomes even more dramatic than the signal. Imposing a lower-limit cut on $\tau_2/\tau_1$ can therefore increase $S/\sqrt{B}$. 

\begin{figure}[thb!]
\begin{center}
\includegraphics[width=0.45\textwidth]{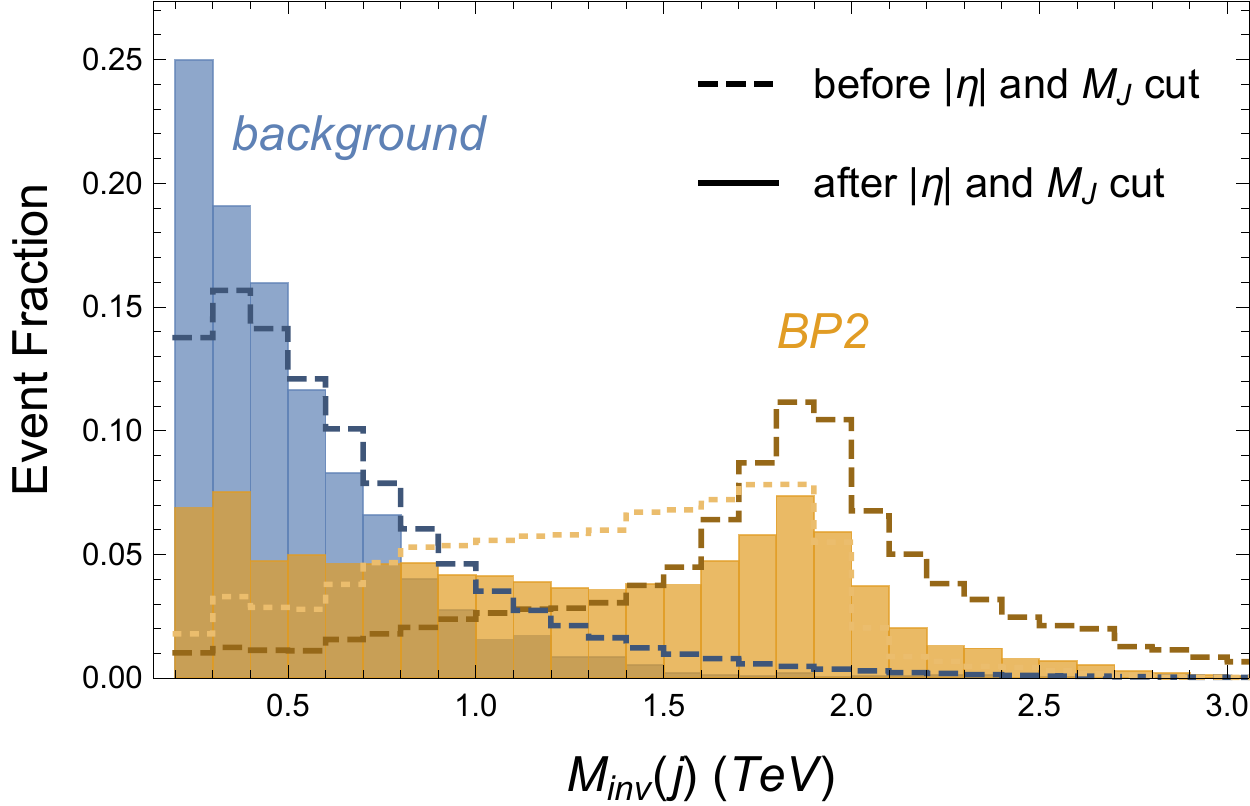}
\hspace{6mm}
\includegraphics[width=0.45\textwidth]{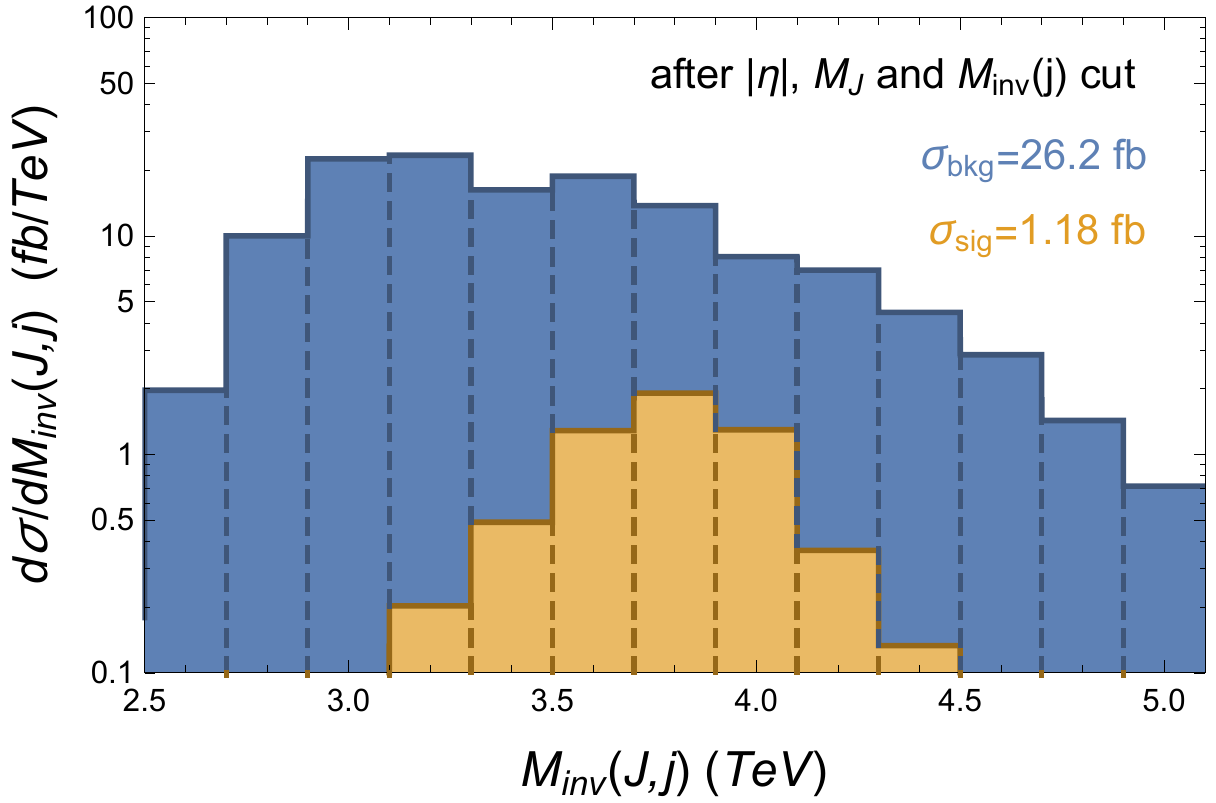}
\vspace*{-0.3cm}
\caption{Left panel: normalized distributions of the invariant mass $M_{\rm inv}(j)$ of all remaining normal jets for the BP2 signal and background before or after the $|\eta|$ and $M_J$ cuts. The yellow dotted line shows the signal distribution if one only includes two leading $p_T$ ordinary jets. Right panel: the invariant mass $M_{\rm inv}(J, j)$ distributions for both the QCD background and the BP2 signal.}
\label{fig:Theta-Gp-mass}
\end{center}
\end{figure}

For the invariant mass $M_{\rm inv}(j)$ of ordinary jets excluding the fat-jet, we show the normalized distributions in the left panel of Fig.~\ref{fig:Theta-Gp-mass}. As one can see, before the $|\eta|$ and $M_J$ cuts, the signal distribution has a nice peak around the $\Theta$ mass. After those two cuts, the spectrum is distorted with a long tail in the small invariant mass region. This is because some jets from $\Theta$ decay may be mis-identified as the fat-jet if the actual $J_{\phi_I}$ does not pass the $|\eta|$ or $M_J$ cuts. For this variable, $M_{\rm inv}(j)$, we have included all ordinary jets passing the basic cuts. In the yellow dotted line, we also show the signal distribution if one only includes two leading $p_T$ ordinary jets and before the $|\eta|$ and $M_J$ cuts. Although it also has a peak at around $M_\Theta$, it does not have a nice bump structure as the yellow dashed line and has a long-tail in lower values due to the final state radiation effect. In the right panel of Fig.~\ref{fig:Theta-Gp-mass}, we show the invariant mass $M_{\rm inv}(J, j)$ distributions after imposing the $|\eta|$, $M_J$ and $M_{\rm inv}(j)$ cuts. For the signal events, they have a peak located at around the true coloron mass of 4 TeV. For the QCD background events, they show a decreasing behavior after around 3 TeV, while the peak at around 3 TeV is mainly due to the cut of requiring $M_{\rm inv}(j)$ close to the $\Theta$ mass of 2 TeV for this BP2. 

\begin{table}[bht!]
\centering
\renewcommand{\arraystretch}{1.2}
\setlength\tabcolsep{0.4em}
\begin{tabular}{c|c|ccc}
\hline\hline
\multirow{2}{*}{Cut-flow} & Background             & \multicolumn{3}{c}{BP2}\\
& $\sigma~(\fb)$ &$\sigma~(\fb)$ & $\epsilon$ & $S/\sqrt{B}$\\
\hline  
$M_J>0.8M_{\phi_I}$\,,  $|\eta| < 2.5$       & $1.60\times 10^4$   	& 6.08	& 81.1\%	& 0.48\\
$M_J>0.8M_{\phi_I}$\,, $\lvert \eta \rvert < 1.0$    							& $1.13\times 10^4$   	& 5.43	& 72.5\%	& 0.51\\
$\lvert M_{\rm inv}(j) - M_{\Theta} \rvert / M_\Theta < 15\%$ 		& 26.6				& 1.18	& 15.7\%	& 2.28\\
$\lvert M_{\rm inv}(J, j) - M_{G^\prime} \rvert / M_{G^\prime} < 15\%$ 	&12.1				& 1.06 	& 14.2\%	& 3.06\\
$\tau_2 / \tau_1 > 0.5$ 									& 2.93				& 0.61	& 8.2\%	& 3.59\\
\hline\hline
\end{tabular}
\caption{Cut flow of signal and background cross sections and acceptances, based on the fat-jet analysis.  Also shown are the signal significances with an integrated luminosity of $100~\ifb$.}
\label{tab:cutflow-BP2}
\end{table}

Using the BP2 as an example, we optimize the cuts for the few kinematic variables in this section together with the $\eta$ cut as
\beqa
&& \lvert \eta \rvert < 1.0 \,, \qquad \tau_2 / \tau_1 > 0.5 \,, \qquad M_J>0.8 \, M_{\phi_I} \,,  \nonumber \\
&&  \lvert M_{\rm inv}(j) - M_{\Theta} \rvert / M_\Theta < 15\%  \,, \qquad  \lvert M_{\rm inv}(J, j) - M_{G^\prime} \rvert/M_{G^\prime}  < 15\%\,.
\label{eq:cuts-BP2}
\eeqa
For the $p_T$ cut used in \eqref{eq:cuts-BP1} of the previous section, it is not efficient to increase the signal significance for the BP2, which is due to the cuts on the variables $M_J$ and $M_{\rm inv}(j)$. So, we just keep the basic cuts of $p_T(j) > 50$~GeV. We also show the cut-flow for the signal BP2 in Table.~\ref{tab:cutflow-BP2}.

\begin{figure}[thb!]
\begin{center}
\includegraphics[width=0.6\textwidth]{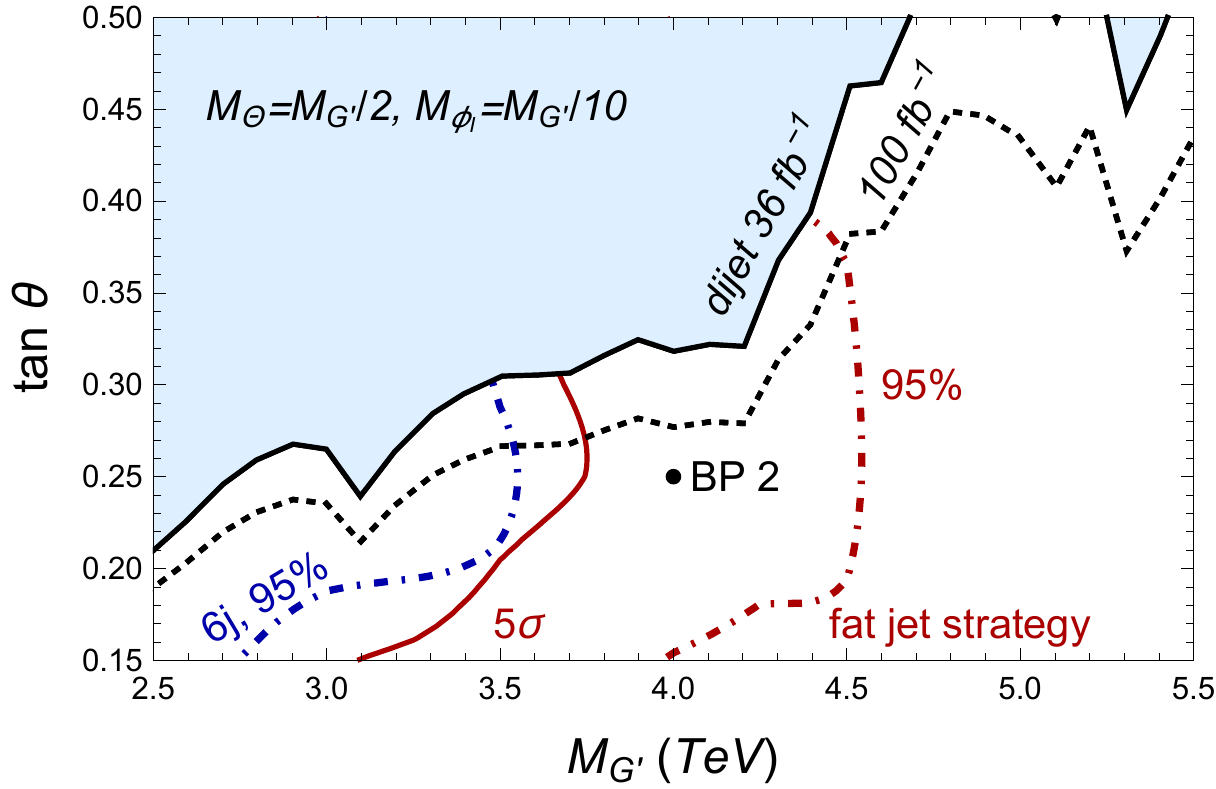}
\vspace*{-0.3cm}
\caption{The same as Fig.~\ref{fig:BP1_contour}, but based on the fat-jet analysis and fixing the ratios of $M_\Theta/M_{G'} = 1/2$ and  $M_{\phi_I}/M_{G'} = 1/10$. The sensitivities for the signal are based on the 13 TeV LHC with 100 fb$^{-1}$. As a comparison, we also show the sensitivities based on the six-jet resonance search strategy in the blue dot-dashed line. }
\label{fig:BP2_contour}
\end{center}
\end{figure}

Using the cuts from \eqref{eq:cuts-BP2}, we scan the parameter space in $\tan{\theta}$ and $M_{G'}$ for the fixed ratios of $M_\Theta/M_{G^\prime} = 1/2$ and $M_{\phi_I} / M_{G^\prime} = 1/10$ and show the $5\sigma$ discovery and 95\% CL exclusion limits in Fig.~\ref{fig:BP2_contour}. In the blue dot-dashed line, we also show the 95\% CL exclusion limit based on the six-jet resonance search strategy in Section~\ref{sec:6-jet}. The comparison between the two sets of exclusion lines shows that the fat-jet based method can probe a larger region of model parameter space. For a wide range of mixing angles, the LHC can discover a coloron via the multi-jet channel up to a mass of around 3.5 TeV and exclude the existence of a coloron at 95\% CL up to a mass of around 4.5 TeV.

%

%==================================
% Conclusion
%==================================
\section{Discussion and conclusions}
\label{sec:conclusion}

In this paper, we have developed two search strategies based on the predicted signatures in the ReCoM. Both strategies could be applied to other models with similar signatures. For instance, one could have the process of $pp \rightarrow Z^{\prime}\rightarrow t^\prime \bar{t}^\prime  \rightarrow W^+ b W^- \bar{b} \rightarrow 4j2b$ with a $Z^\prime$ decaying into some top-partner $t^\prime$. The $Z^\prime$ behaves as a six-jet resonance with the hadronic $W$'s as possible fat-jets. Although there are non-resonance searches based on multi-jet final states~\cite{Sirunyan:2017anm,Sirunyan:2018xwt,Aaboud:2018lpl} at the LHC, looking for additional ``bumps" that are composed a large number of jets has a great discovery potential. 

We have concentrated on the prompt decays of new particles in this paper. As pointed out in Ref.~\cite{Bai:2018jsr}, the color-singlet scalar $\phi_I$ may have a long decay lifetime if its mass is below around 500 GeV, with the main decay channel as two electroweak gauge bosons. It is feasible that the discovery potential is even higher than both results in Fig.~\ref{fig:BP1_contour} and \ref{fig:BP2_contour}, if the displaced vertex can be identified inside the detector. In the ReCoM and because of a $U(1)$ symmetry, the $\phi_I$ could be a long-lived PNGB. Although its direct couplings to the SM particles are suppressed, it can be produced via decays of other heavy new particles like $G'$ considered in this paper. A similar situation may happen in other beyond-SM models. For instance, the top-partner $t^\prime$ can decay into a top quark plus a long-lived PNGB~\cite{Bizot:2018tds}. Searching for this type of signatures with long-lived particles at the LHC can explore a wide range of new physics models. 

In summary, we have performed a detailed collider study for one type of coloron signatures at the LHC. The coloron can behave as a six-jet resonance from its cascade decays of $G^\prime \rightarrow (\Theta \rightarrow gg) (\phi_I \rightarrow q\overline{q}g g )$. For the case with a non-boosted $\phi_I$, we have found that performing a bump search for a six-jet resonance can cover a wide and new region of coloron parameter space. For the mixing angle below 0.35 and above the perturbative value of 0.15, a coloron with a mass from 3.3 TeV to 4.0 TeV can have a $5\sigma$ discovery chance, while at 95\% CL its mass can be constrained up to 5 TeV. For the second case with a boosted $\phi_I$, a fat-jet-based strategy is developed with the $5\sigma$ discovery limit up to 3.5 TeV and the 95\% CL constraint limit up to 4.5 TeV for the coloron mass. The large list of new signatures for the coloron in Ref.~\cite{Bai:2018jsr} and the specific signatures studied in this paper serve as examples for the great discovery potential of new physics based on future LHC data.

\vspace{3mm}
%----------------------------------------------------------------
% Acknowledgements
%----------------------------------------------------------------
\subsubsection*{Acknowledgements}
We thank Bogdan A. Dobrescu for discussion.  
The work of YB and SL is supported by the U. S. Department of Energy under the contract DE-SC0017647. The work of QFX is supported by the China Scholarship Council with Grant No. 201704910714. 

\newpage
%----------------------------------------------------------------
% Appendix
%----------------------------------------------------------------
\begin{appendix}

%-------------------------------------------
\section{$\tau_4/\tau_3$ distribution for the fat $\phi_I$ jet}
\label{appendix:tau43}
\begin{figure}[hb!]
\begin{center}
\includegraphics[width=0.5\textwidth]{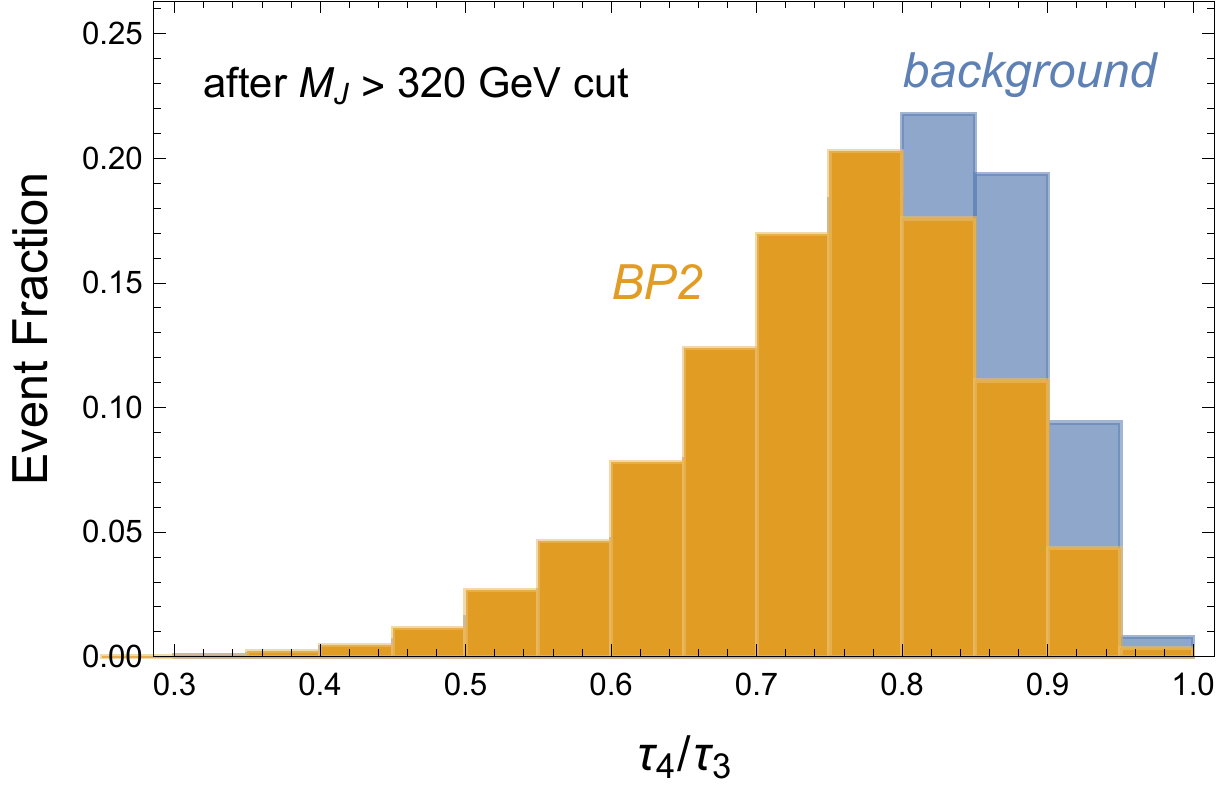}
\vspace*{-0.3cm}
\caption{Normalized $\tau_4/\tau_3$ distributions of the most massive jet for the signal and background after a jet-mass cut.}
\label{fig:tau43}
\end{center}
\end{figure}

Because of the decay of $\phi_I \rightarrow gg q\bar{q}$, a boosted $\phi_I$ is anticipated to behave as a four-prong fat-jet. Naively, one should anticipate that the N-subjettiness variable, $\tau_4/\tau_3$, could be useful to separate signal from background. However, the four-body decays of $\phi_I$ is through off-shell $G'$ and $\Theta$. As a result, the four partons in the decay products are more hierarchic in energy rather than more democratic, which downgrades its four-prong feature. In Fig.~\ref{fig:tau43}, we show the distributions of $\tau_4/\tau_3$ for both signal and background, which shows the similarity of the two distributions.

\end{appendix}

%----------------------------------------------------------------
% References
%----------------------------------------------------------------
%\bibliographystyle{JHEP}
%\bibliography{Coloron}
\providecommand{\href}[2]{#2}\begingroup\raggedright\endgroup

\end{document}